\documentstyle[12pt]{article}

\begin{document}
\centerline{\large\bf Resonance propagation in heavy-ion scattering}\par
\vskip 0.5cm
\centerline{Bijoy Kundu$^a$\footnote{email: bijoyk@iitk.ac.in} and 
B. K. Jain$^b$\footnote{email: bkjain@magnum.barc.ernet.in}
}  
\vskip 0.5cm
\centerline{$^a$Department of Physics, IIT Kanpur, Kanpur-208 016, India}
\centerline{$^b$Nuclear Physics Division, Bhabha Atomic Research Centre,}\par
\centerline{Mumbai - 400 085, India}

\vskip 1.0cm
PACS number(s): 25.75.+r;24.30.-v;24.50.+g;24.10.-i 
\vskip 2.0 cm
\centerline{\bf Abstract}
The formalism developed by Jain and Kundu
for the propagation of a resonance in the nuclear medium has been 
modified to the
case of heavy-ion collisions at relativistic energies. 
The formalism includes coherently the contribution to the observed 
di-lepton production from the decay of the resonance inside as well as 
outside the nuclear medium. The calculated results are presented for 
the $\rho $-meson production.
It is observed that, in general, the shape, magnitude
and peak position of the coherently summed invariant mass distribution is 
much different from that obtained by summing the inside and outside 
contributions incoherently.
Therefore, while inferring the modification of hadron properties 
produced in heavy-ion collisions from experiments, it is important
that in theoretical calculations one includes the decay of the
propagating hadron from inside and outside the 
heavy-ion system coherently. We also find that the mass distribution 
is sensitive to the amount of the medium modification of the 
$\rho $-meson.
\newpage
\section {Introduction}

The modification of hadron masses in the nuclear medium, either
in nucleon-nucleus or nucleus-nucleus collisions, is an issue
of much interest currently \cite{Bores}. Since in QCD the hadrons
are excitations of the vacuum, it is natural that these excitations can 
get affected by the proximity of other hadrons. Experimentally the masses
of unstable hadrons are explored by producing them in nuclear reactions
and then measuring the invariant masses of their leptonic decay products
\cite{Aga}. The medium modification of the unstable hadron from these 
data is, normally, inferred by adding incoherently the decay of the hadron
inside and outside the nuclear medium.  

Recently Jain et. al \cite{Jain} have developed a formalism f the  
propagation of resonances produced in proton-nucleus collisions. 
This formalism incorporates the interaction of the resonance with the  
nuclear medium and the coherent contribution from its decay inside as well
as outside the nucleus to the measured decay products.
In this paper we modify this formalism
and apply it to the propagation and decay 
of unstable 
hadrons in heavy-ion collisions,
wherein two heavy nuclei collide at relativistic energies. 
We assume that the resonance is produced 
when two nucleons, one in the target $B$ and another in the projectile $A$, 
collide. From then on the resonance moves along the z-axis, which
is taken along the beam direction. We do not worry about the state of the
$(A+B)$ system.  
The resonance coordinate is with respect to the c.m of the
(A+B) system. The propagating resonance can decay either inside or outside the
$(A+B)$ system and both these contributions
add coherently. The formalism developed here could be 
applied in general
to any propagating resonance produced in nucleus-nucleus collisions.
However,  we study, specifically, 
the invariant mass spectra of a $\rho$ resonance,
traversing at different velocities, produced in $Pb+Pb$ and $S+Au$
collisions. We also investigate the decay probability
of the $\rho$ resonance inside the nuclear medium as a function of 
its speed of propagation.

The general findings of this paper are that the probability of the
$\rho$ resonance decaying inside the $A+B$ system is large. 
Still, the summed cross-section
(i.e. its magnitude, shape and the peak position), especially at higher
speeds where the contribution from the outside increases, is determined
equally by the contributions from inside as well as  
outside. We also find that the calculated cross sections differ 
a lot if they are calculated by adding coherently or incoherently the 
contribution from inside and outside the nucleus. So, while infering
the modification of hadron masses produced in nucleus-nucleus collisions
at relativistic energies from experiments, it is important that in theoretical
framework one includes the contribution of the decaying hadron from the
the inside as well as outside the $(A+B)$ system coherently.

\section {Formalism}

Let us suppose that two heavy nuclei, one the projectile $A$ and other the
target $B$, collide at relativistic energies. We assume that one nucleon
in the projectile and one in the target collide and 
a resonance $R$ is produced at the collision point. 
This resonance then moves along the beam direction, which
is taken as the z-axis. Essentially, we study the reaction,
\begin{eqnarray}
A +B \rightarrow R + (A+B),
\end{eqnarray}
where, $A$ is the projectile and $B$ is the target nucleus. $R$ is the 
resonance produced in the nuclear collisions which decays into
say $x$ and $y$, i.e.
\begin{eqnarray}
R \rightarrow x + y .
\end{eqnarray}
Denoting by ${\bf r}$  the relative coordinate
between the target and the projectile, and by 
${\bf (r_{B}, r_{A})}$ the intrinsic coordinates of
the target and projectile nucleons, respectively, 
for the resonance coordinate we write, 
\begin{eqnarray}
& &{\bf r_R} = {\bf r_A} + \frac {B} {A+B} {\bf r}.
\end{eqnarray}
With this defination, for an inclusive situation where the state of the 
(A+B) system is not identified, the ratio of the cross sections for the 
resonance production in AB and NN collisions can be written as,
\begin{eqnarray}
\frac {\Delta \sigma _R^{AB}}{\Delta \sigma _R^{NN}} =[K.F.] \int d {\bf b}
\int dz \int d {\bf r_A} \rho_A ({\bf r_A}) \rho_B ({\bf r + r_A})
|G({\bf r_R;k_R,\mu})|^2,
\label{sigma}
\end{eqnarray}
where $\rho_x$ are the nuclear densities. [K.F.] is the relevant kinematic
factor.
The function $G({\bf r_R;k_R,\mu})$ physically gives the probability amplitude 
for finding the decay products of the resonance in the detector with 
the total momentum 
\begin{eqnarray}
{\bf k_x+k_y = K_R}
\end{eqnarray}
and the invariant mass $\mu $ if the resonance is produced at a point
${\bf r_R}$ in the nucleus (for details see \cite {Jain}).
In terms of the resonance propagator $G({\bf r^{'}_R,r_R})$, 
function  $G({\bf r_R;k_R,\mu})$ is defined as 
\begin{eqnarray}
G({\bf r_R;k_R,\mu}) = \int d{\bf r^{'}_R } exp(-i {\bf k_R.r^{'}_R})({\bf r^{'}_R,r_R}),
\end{eqnarray}
where $G({\bf r^{'}_R,r_R})$ satisfies
\begin{eqnarray}
[\nabla^2 + E^2 -m^2_R +i \Gamma_R m_R -\Pi_R] G({\bf r^{'}_R,r_R})= 
\delta ({\bf r^{'}_R-r_R}).
\end{eqnarray}
Here $\Pi_R$ is the self energy of the resonance in the medium and 
$\Gamma_R$ is its free space decay width.

In the eikonal approximation we can write,
\begin{eqnarray}
G({\bf r_R;k_R,\mu})=exp(-i {\bf k_R.r_R})\phi_R({\bf r_R;k_R,\mu}),
\end{eqnarray}
where $\phi_R$ is a slowly varying modulating function. With this, and 
using the eqns. (6, 7), $\phi_R$ approximately 
works out to
\begin{eqnarray}
\phi({\bf r_R;k_R,\mu}) &=& \frac {1} {2ik_R} \int dz^{'}_R exp[\frac {1}{2ik_R}
(\mu^2 -m^2_R+i\Gamma_R m_R)(z_R-z^{'}_R)] \nonumber \\
& &\times exp[\frac {-i}{v_R} \int ^{z'_R}_{z_R}
V_R (b_R,z{''}_R) d z^{''}_R].
\end{eqnarray}
Here we have written 
\begin{eqnarray}
\Pi_R= 2 E_R V_R,
\end{eqnarray}
where $V_R$ is the optical potential of R in the 
nuclear medium. In general, it is complex. Its real part, as we shall see 
later, is related to the mass shift of the resonance and the imaginary 
part gives the collision broadening of the resonance in the medium. 

For a nucleus with a sharp surface, function $\phi({\bf r_R;k_R,\mu})$ 
splits into a sum of two terms, one corresponding to the decay of the 
resonance inside the nucleus and another to the decay outside the nucleus,
i.e.
\begin{eqnarray}
\phi({\bf r_R;k_R,\mu}) = \phi _{in} ({\bf r_R}) + \phi_{out} ({\bf r_R})
\end{eqnarray}
with
\begin{eqnarray}
\phi_{in}({\bf r_R}) = \frac{1}{2ik_R} \int_{z_R}^{\sqrt (R^2-b^2)} d z'_R 
\phi_R
({\bf b_R}; z_R,z'_R),
\end{eqnarray}
and
\begin{eqnarray}
\phi_{out}({\bf r_R}) = \frac{1}{2ik_R} \int_{\sqrt (R^2-b^2)}^{\infty} d z'_R 
\phi_R
({\bf b_R}; z_R,z'_R).
\end{eqnarray}
Here
\begin{eqnarray}
\phi_R({\bf b_R}; z_R,z'_R) &=& exp[\frac {1}{2ik_R}
(\mu^2 -m^2_R+i\Gamma_R m_R)(z_R-z^{'}_R)] \nonumber \\
& & \times exp[\frac {-i}{v_R} 
\int ^{z'_R}_{z_R}
V_R (b_R,z{''}_R) d z^{''}_R].
\end{eqnarray}
After a little bit of manipulations, the final expressions for 
$\phi_{in}$ and $\phi_{out}$ work out to ,
\begin{eqnarray}
\phi_{in}({\bf r_R}; k_R,\mu) = \frac {G_0^*} {2 m_R} [1 - exp(\frac {i} 
{v_R G_0^*} [L(b_R) - z_R])],
\label{phiin}
\end{eqnarray}
\begin{eqnarray}
\phi_{out}({\bf r_R}; k_R,\mu) = \frac {G_0} {m_R} [ exp(\frac {i} 
{v_R G_0^*} [L(b_R) - z_R])].
\label{phiout}
\end{eqnarray}
where $v_R$ is the speed of the resonance and $
L (= \sqrt{(R^2 -b R)})$ is the length from the production point to the
surface of the nucleus.
$G_{0}$ and $G_{0}^{*}$
in above equations, \ref {phiin} and \ref {phiout},  are the
free and the in-medium resonance propagators. Their forms are 
\begin{eqnarray}
G_0= \frac{2 m_R} {\mu^2 - m^2_R +i \Gamma _R m_R},
\end{eqnarray}
\begin{eqnarray}
G_0^*= \frac{2 m_R} {\mu^2 - m^{*2}_R +i \Gamma _R^* m_R},
\end{eqnarray}
with
\begin{equation}
E_R^2=k_R^2+\mu ^2,
\end{equation}
\begin{eqnarray}
m_R^{*2} &=&  m^2 _R + \Pi_R 
\nonumber \\
&=& m^2_R + 2 E_R U_R. 
\end{eqnarray}
or
\begin{eqnarray}
m_R^*\approx&  m_R + \frac{E_R}{m_R} U_R .
\end{eqnarray}
\begin{eqnarray}
\Gamma ^* _R &=& \Gamma_R -2 \frac{E_R} {m_R} W_R 
\nonumber \\
&=& \Gamma_R + \frac{E_R}{2m_R} |2 W_R| 
\nonumber \\
&=& \Gamma_R + \frac{E_R}{2m_R} \Gamma _{collision}.
\end{eqnarray}
In the above we have written,
\begin{eqnarray}
V_R= U_R + i W_R .
\end{eqnarray}
For an estimate of this potential we use the high energy ansatz, and write
\begin{eqnarray}
U_R = -\alpha [\frac{1}{2} v_R \sigma _ T^{RN} \rho_0]
\end{eqnarray}
and 
\begin{eqnarray}
W_R= -[\frac{1}{2} v_R \sigma _ T^{RN} \rho_0],
\end{eqnarray}
where $\alpha$ is the ratio of the real to the imaginary part of the 
elementary RN scattering amplitude and $\sigma _ T^{RN}$ is the total
cross section for it. $\rho_0$ is the typical nuclear density. Taking
$\sigma_{T}^{\rho N}$ = 38 mb \cite{Gott},
 $\alpha$=0.7 and $\rho_0$=0.17 fm$^{-3}$, we get the optical potential 
 for the $\rho$
meson to be, $\frac {V}{v_R}$=-(45+i64) MeV. For higher nuclear densities, 
these values would be higher.

\section {Results and Discussion}
Examining above eqns. we see that the propagation of the resonance in 
the nucleus depends upon its speed ($v_R$), the length of the nuclear
medium, the free decay width of the resonance and its self energy in the 
medium. We study the decay of rho-meson in the nucleus, 
and present results for two values of $v_R$ and two sets of 
nuclear systems. The free width of the rho-meson is taken equal to 150 MeV.
For nuclei we have taken the realistic densities from \cite{Vries}.
 
Denoting the ratio $\frac {1}{[K.F.]}
\frac {\Delta \sigma _R^{AB}}{\Delta \sigma _R^{NN}}$ 
in equation \ref{sigma} as $|\Phi|^2$, we plot it as a function of
the invariant mass, $\mu$, of the decay products $x$ and $y$ of the $\rho$
meson. Figures 1-4 show the invariant mass spectra
of the $\rho$ meson in $Pb+Pb$ and $S+Au$ collisions at rho velocities of
$0.6c$ and $0.9c$. Here $c$ is the speed of light.
 The solid curve in all the 
figs. gives the total summed cross-section
(coherently) and the dashed curve gives the same added incoherently.
The individual contributions corresponding to the inside and the outside decay
are given by the dotted and dense-dotted curves respectively. We observe
two things. One is that the coherent and the incoherent cross-sections
are different and second, this difference increases with the 
increase in the rho-meson speed. At 0.6c speed, while the coherent and 
incoherent curves differ only in the peak cross sections, at 0.9c speed 
their shape and peak cross sections both are different. 
We also observe that the 
difference is larger for the smaller system like S on Au.

Compared to free $\rho$-meson, walso observe that in $Pb+Pb$ collisions 
at $0.6 c$ rho-meson speed the peak positions
are not really effected so as to be detected experimentally. The width
of the summed cross-section, of course, broadens to about twice the 
free width, which
may be attributed to the collisions of the propagating $\rho$
meson with the nucleons in the nucleus. The width of the inside decay
is almost same as that of the summed cross-section, whereas the outside decay
has the free width of about 150 MeV. 

At  $0.9 c$ speed (fig. 2)
the inside decay spectrum shows a shift towards lower mass by
about 70 MeV. But this shift is not directly visible in the summed cross 
section because the contribution from the outside decay also increases, 
which does not have any shift. The shift in the inside contribution 
is perceptible only through the skewing of the shape of the summed cross 
section.  The width of the inside spectrum is also broadened 
considerably and so is the coherent cross-section

Observations for the $S+Au$ collisions are similar.

To give a quantitative measure of the $\rho $ decay probability 
inside and outside the nucleus,
in fig. 5 we have the fractional outside decay probability 
(eqs. 15, 16) as a function of the
increasing rho meson velocity. The results are shown for the $Pb+Pb$ and 
$S+AU$ systems 
integrated over the   kinematically allowed invariant mass.
We observe that as the rho velocity increases the outside decay probability
increases, and can go upto 10$\%$ and 25$\%$ respectively in two cases.

Finally, in figure 6 we explore the sensitivity of the cross
section to the amount of mass modification of the rho-meson. For this we 
use eq. (21), which relates the mass-shift $(m^*_R-m_R)$ to the 
optical potential $U_R$. We show the calculated cross sections for mass 
shift equal to 50, 100 and 150 MeV for $Pb+Pb$ system at $0.9c$ rho speed.
We see that the curves for different masses are very different in shape 
as well as magnitude. With the increase in mass shift, in fact, the curve 
eventually becomes quite flat in the region of lower 
invariant mass. This resembles 
somewhat the $\rho $ invariant mass spectrum seen in heavy-ion experiments.  

\section {Conclusions}

Modifying the formalism of Jain and Kundu \cite {Jain}, to the
case of heavy-ion scattering, we find that the summed cross-sections
(i.e. its magnitude, shape and the peak position), especially at higher
speeds where the contribution from the outside decay of the $(A+B)$
system increases, is determined by the contributions from the
outside as well as inside of the nucleus. Therfore, while inferring the 
modification of
hadron properties produced in heavy-ion collisions from experiments,
it isportant that in theoretical calculations one should include
the contribution from the decay of hadrons from the inside as well as
outside the heavy-ion system coherently. We also find that the shape
of the invariant rho mass distribution is very sensitive to its mass 
modification in the medium. For large mass shifts ($\sim $100 MeV and beyond)
the mass distribution starts becoming very broad, as indicated in 
heavy-ion experiments.

\newpage
\section{Figure Captions}
\begin{enumerate}
\item The invariant mass spectra of the $\rho$ meson produced
in Pb+Pb collisions. The solid curve is obtained after adding
the inside and the ouside decay coherently. The dashed curve is obted
incorently. The dotted curve is the inside decay amplitude and the
dense dot curve is the outside decay amplitude. The resonance speed
is 0.6c.
\item All curves have  same meaning as in fig 1. The resonance
speed is 0.9c.
\item All curves have  same meaning as in fig. 1. The colliding
nuclear system is S+Au and the resonance speed is 0.6c.
\item Same as in fig. 3 . The resonance speed is 0.9c.
\item The outside decay probability as a function of the
resonance velocity. The solid curve is for S+Au and the dashed curve is for
Pb+Pb nuclei.
\item Sensitivity of the cross-section to the amount
of mass modification of the rho-meson. The calculations are presented
for Pb+Pb system at v=0.9c of the rho-meson.
The curves a, b and c correspond to the mass shifts 50, 100 and 150 MeV
respectively.
\end{enumerate}
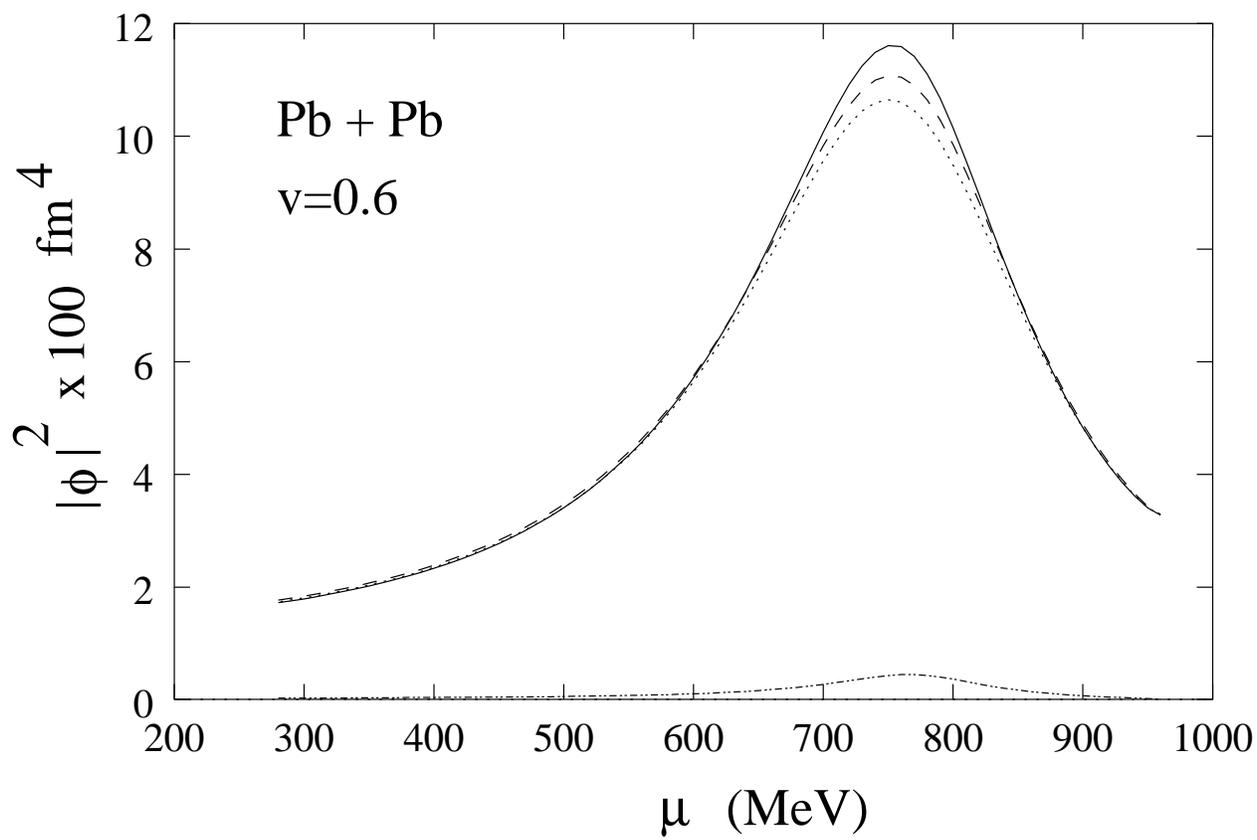
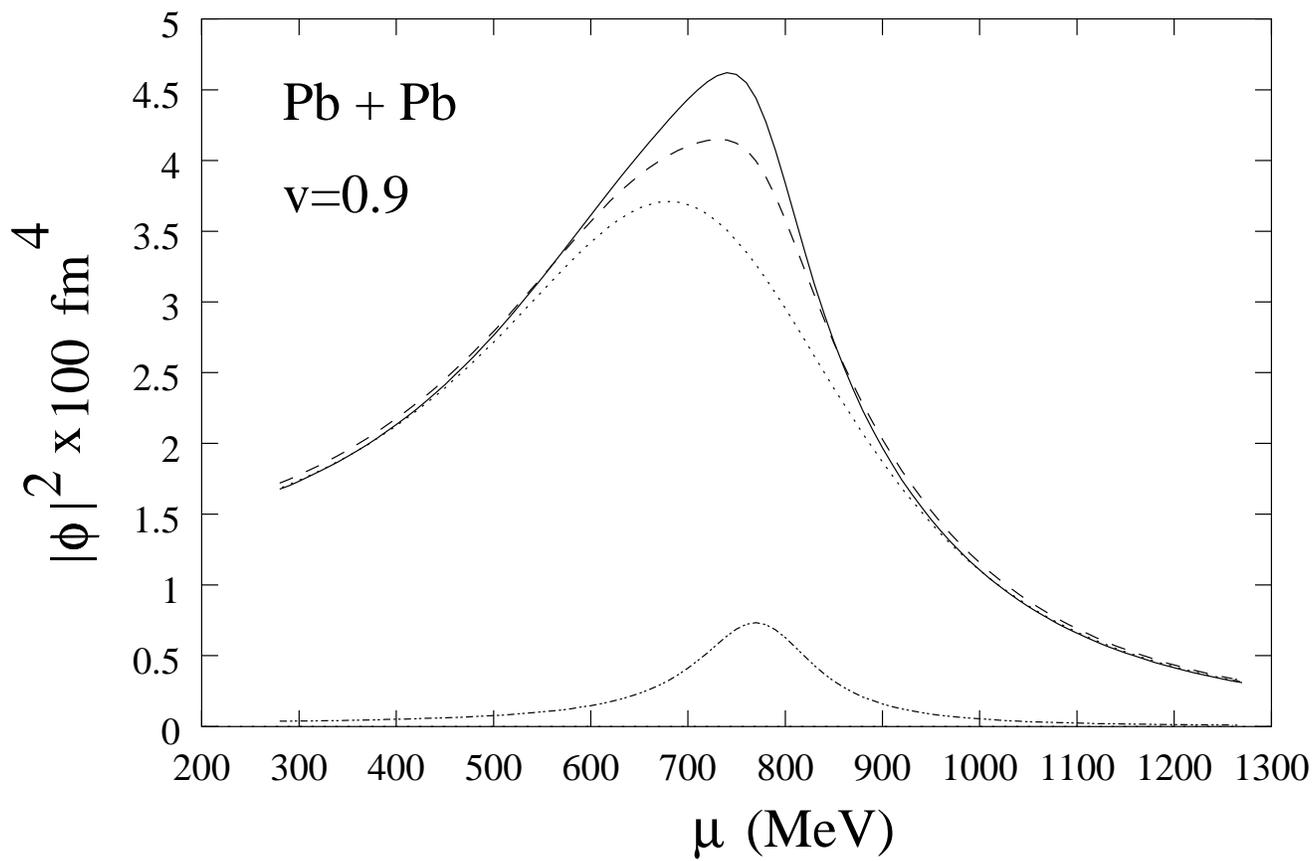
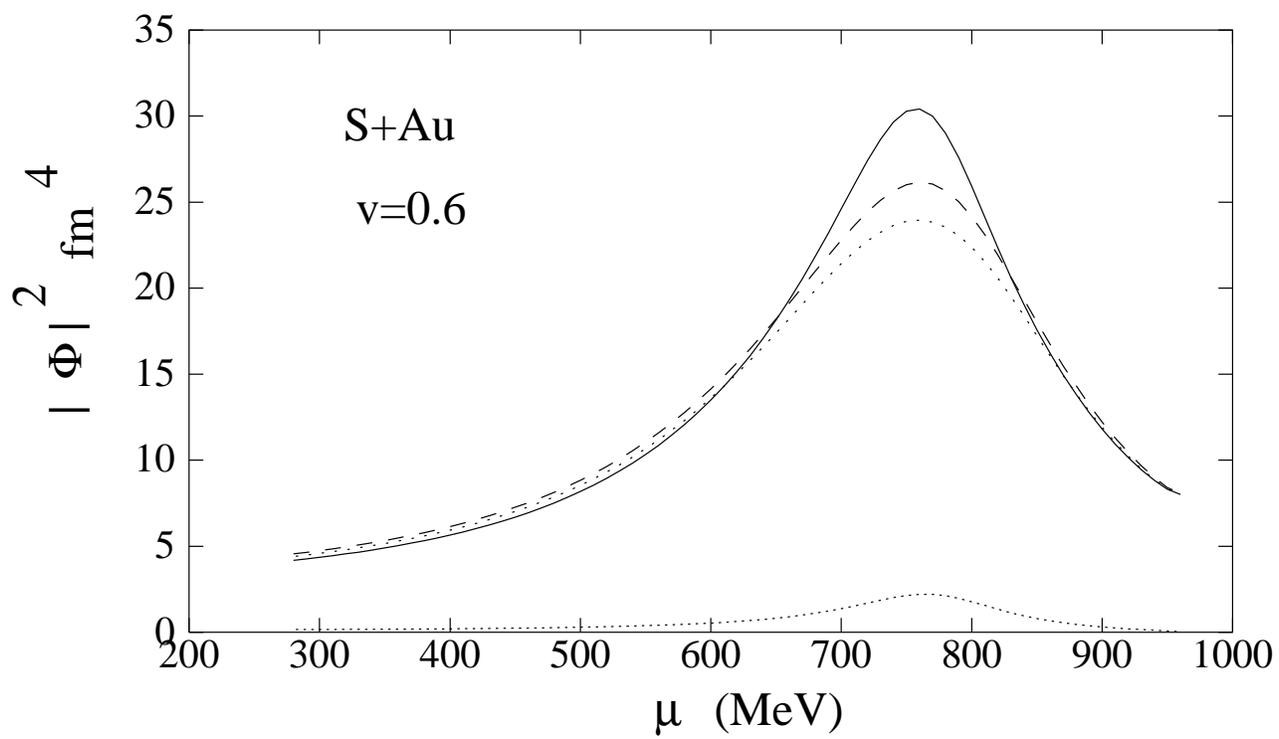
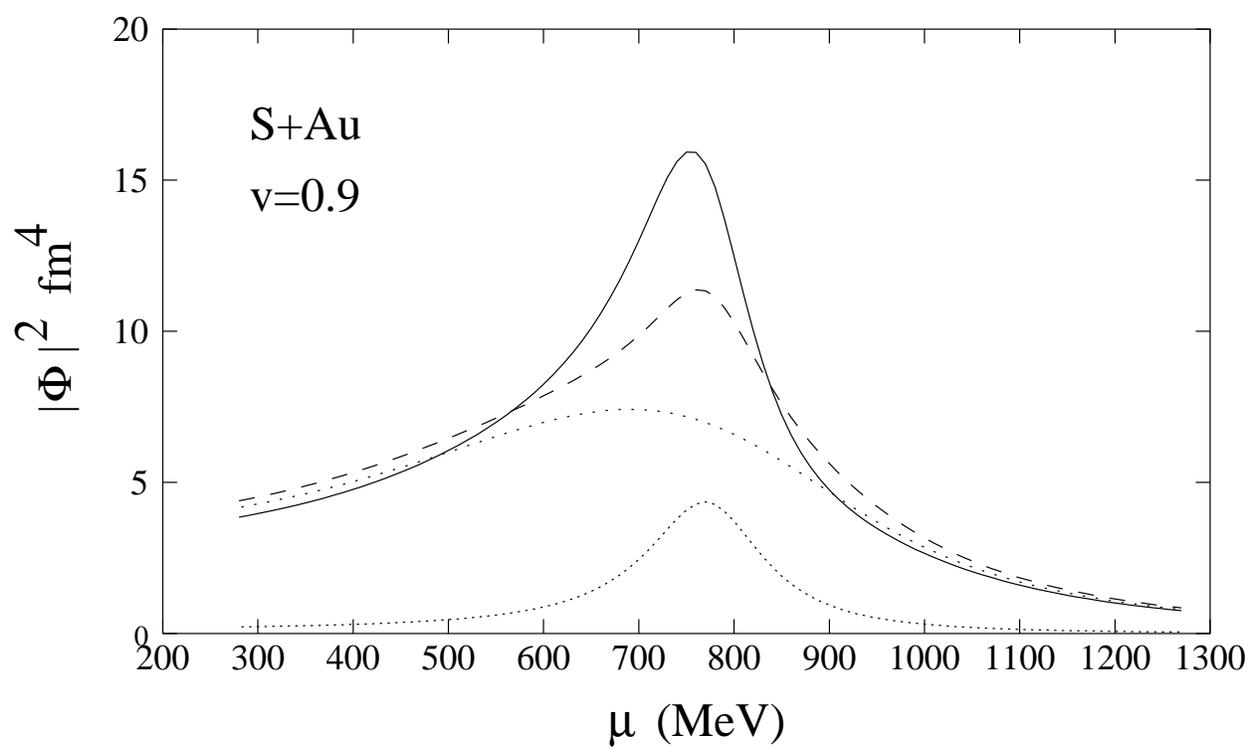
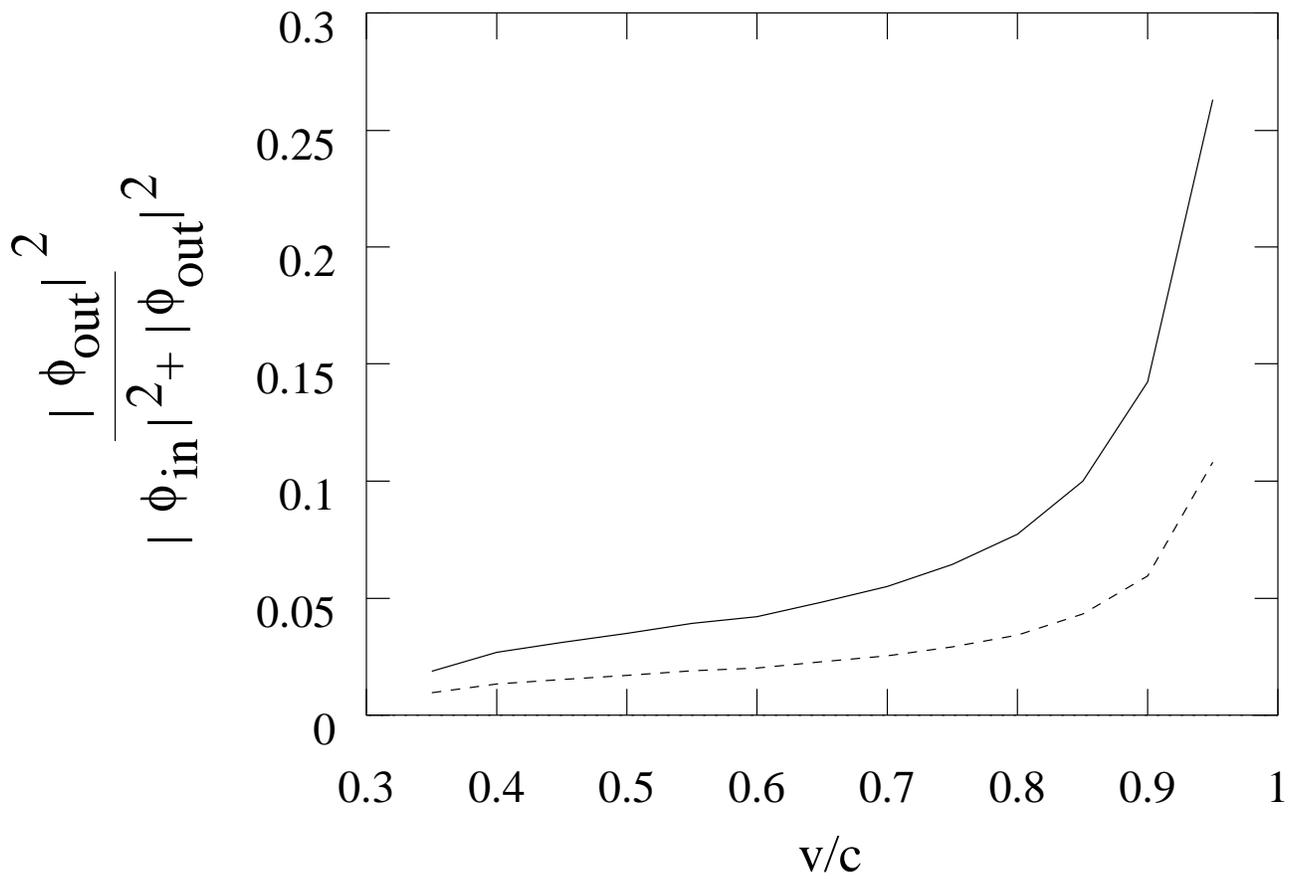
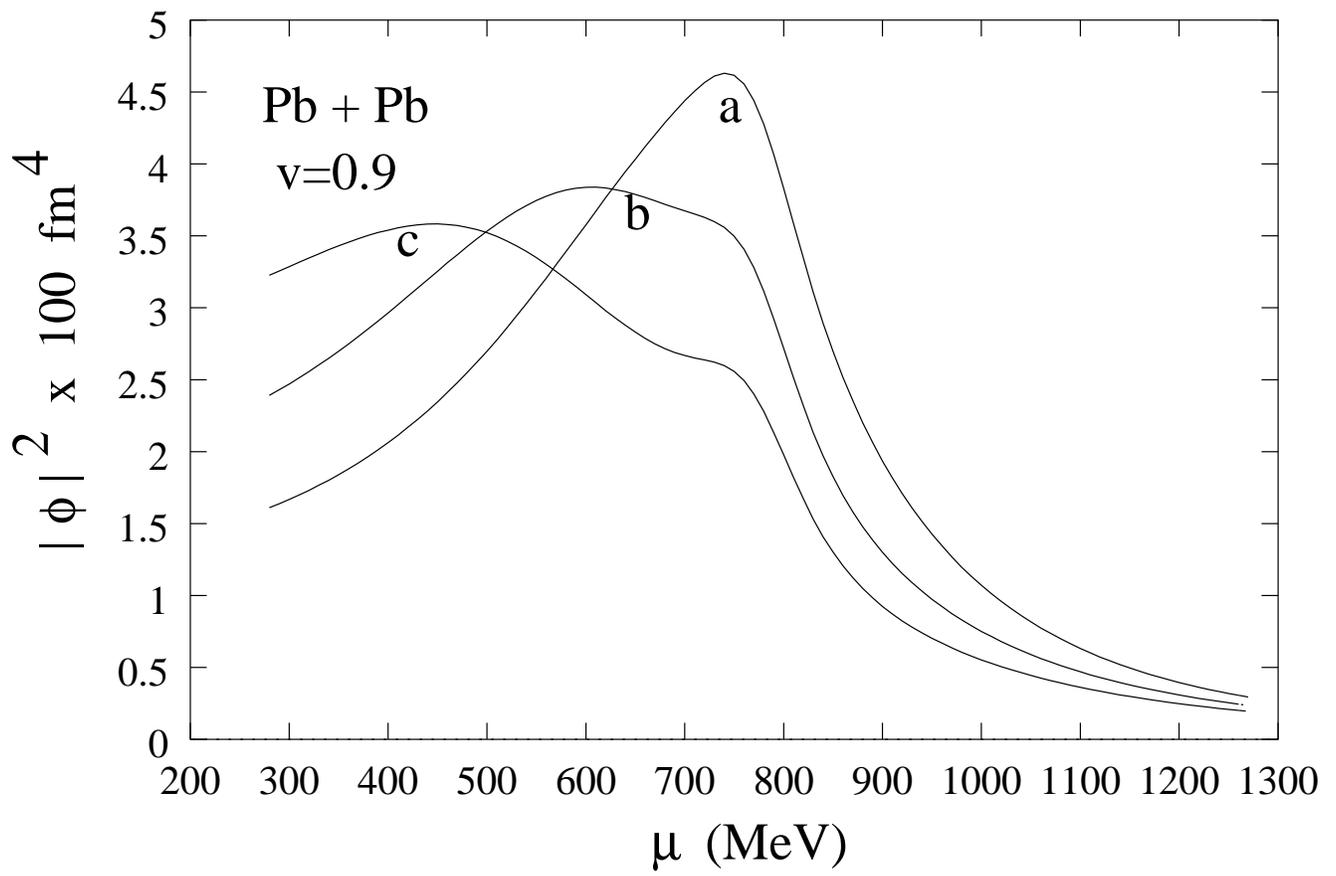
\end{document}